\documentclass[11pt]{article}
\usepackage{moriond,epsfig}
\usepackage{graphics,amssymb,float}
\usepackage{amsmath,mathbbol}
\usepackage{amsfonts}

\def\be{\begin{equation}}
\def\ee{\end{equation}}
\def\bea{\begin{eqnarray}}
\def\eea{\end{eqnarray}}

\begin{document}
\begin{flushright}
PCCF RI 1103
\end{flushright}
\vspace*{4cm}
\title{The Next-to-Minimal Supersymmetric Standard Model: an overview}

\author{Ana M. Teixeira}

\address{Laboratoire de Physique Corpusculaire, CNRS/IN2P3 -- UMR 6533,\\ 
Campus des C\'ezeaux, 24 Av. des Landais, F-63171 Aubi\`ere Cedex, France}

\maketitle\abstracts{
We review the most important aspects of the NMSSM,
discussing the impact of the NMSSM on low-energy observables, for
dark matter, as well as NMSSM specific signatures at colliders. We
also briefly consider constrained realisations of the NMSSM.}

\section{Supersymmetric extensions of the Standard Model}

Among the many extensions of the Standard Model (SM) which  aim
at solving or easing its observational and theoretical shortcomings,
supersymmetry (SUSY) is one of the most appealing possibilities. 
SUSY extensions of the SM offer a potential solution to the hierarchy
problem, allow for radiative spontaneous
electroweak (EW) symmetry breaking, and provide a possible link
between the EW scale and the scale of soft-supersymmetry breaking
($M_\text{SUSY}$). SUSY models are further motivated by an automatic
unification of the running gauge coupling constants under simple SU(5) or
SO(10) grand unified (GUT) models, at a scale $10^{16} \text{ GeV}\lesssim
M_\text{GUT} \lesssim 10^{17}$ GeV. 
If R-parity is conserved, the lightest SUSY particle (LSP) is stable;
if neutral and colourless, it can be a candidate to 
explain the observed dark matter (DM) relic density of the Universe.

The minimal supersymmetric extension of the SM (MSSM) is defined by
the following superpotential and supersymmetry soft-breaking Lagrangian  
\begin{equation}\label{eq:MSSM:W}
\mathcal{W}= Y_u \, \hat H_u\, \hat Q \, \hat u +Y_d \, \hat H_d\, \hat Q \,
\hat d +Y_e \, \hat H_d\, \hat L \, \hat e + \mu \,\hat
    H_u \hat H_d\,,
\end{equation}
\begin{equation}\label{eq:MSSM:Lsoft}
-\mathcal{L}_\text{soft}= m_{H_u}^2\,H_u^*H_u +    
m_{H_d}^2 H_d^* H_d + (M_i \psi_i \psi_i +A_F Y_F H_i \tilde F \tilde F^*+
B \mu H_u H_d + \text{\small H.c.})+ ...
\end{equation}

\noindent
Other than squarks, sleptons and gluinos, the spectrum contains 
2 charginos and 4 neutralinos, arising from  the mixing of  
electroweak gauginos with the charged and neutral
fermion components of the two Higgs superfields, $\hat H_d$ and $\hat
H_u$.
The Higgs sector is composed of 2 neutral
scalars ($H_i$), one pseudoscalar $A$, and a pair of charged states
$H^\pm$.  

\medskip
Despite its many appealing features, the MSSM suffers from
phenomenological problems; among them, and deeply related  
to the Higgs sector, is the so-called ``$\mu$-problem''~\cite{Kim:1983dt}. 
The latter arises from the presence of a non-vanishing 
dimensionful term in the MSSM superpotential of
Eq.~(\ref{eq:MSSM:W}), for which there are only two ``natural'' values:
either 0, or then the typical scale at which the model is defined 
($\sim M_\text{GUT,\,Planck}$). However, and as we briefly discuss, 
neither possibility is viable.
The non-observation of charginos at LEP puts
a limit on their mass ($m_{\chi^\pm_1} \gtrsim 103$ GeV), and
hence a lower bound on the SUSY conserving mass term, $\mu
\tilde h_u \tilde h_d$, $|\mu| \gtrsim 100$ GeV. In any case, in order
to ensure that the neutral components of both Higgs scalars develop
non-vanishing vacuum expectation values (VEVs), $\mu \neq 0$. 
Moreover, a correct EW symmetry
breaking implies that the SUSY conserving $\mu$ 
term cannot be excessively large: 
the $\mu$-induced mass squared for $H_u$ and $H_d$ (always positive)
must not dominate over the negative soft breaking masses, which
further precludes $\mu \sim M_\text{GUT,\,Planck}$. Everything taken 
into account, $\mu$ must 
be of order of the soft SUSY breaking scale, $|\mu| \sim 
\mathcal{O}(M_\text{SUSY})$, which is
a very unnatural scenario.

An elegant and yet simple way to solve this problem consists in the
addition of a superfield to the MSSM content, and in taking
a scale-invariant superpotential where only trilinear dimensionless
couplings are present. The required non-vanishing bilinear mass term
for the Higgs can then be effectively generated from the VEV of the new
scalar field (necessarily a singlet since the $\mu$-parameter is gauge
invariant): $\mu^\text{eff}= \lambda \langle S \rangle$.
This is the so-called Next-to-Minimal supersymmetric
standard model (for a recent review, see~\cite{Ellwanger:2009dp}). 

\section{The Next-to-Minimal Supersymmetric Standard Model}
In its simplest form, the Next-to-Minimal supersymmetric
standard model (NMSSM) is described by the superpotential
\begin{equation}\label{eq:NMSSM:W}
\mathcal{W}^\text{NMSSM}= Y_u \, \hat H_u\, \hat Q \, \hat u +Y_d 
\, \hat H_d\, \hat Q \,
\hat d +Y_e \, \hat H_d\, \hat L \, \hat e + \lambda\, \hat S \,\hat
    H_u \hat H_d+\frac{1}{3}\kappa\,\hat S^3\,.
\end{equation}
In the soft breaking Lagrangian, the $B\mu$ term is replaced by
trilinear couplings, $A_\lambda$ and $A_\kappa$, 
and there is an additional soft breaking
mass for the scalar, $m_S^2$. Phenomenologically viable values of
$\mu_\text{eff}$ can be easily obtained with negative soft SUSY
breaking mass squared (and trilinear couplings) for the singlet. 
It is also important to stress that in this case, 
all the fermions belonging to a chiral superfield will
have a supersymmetry conserving mass term in the Lagrangian arising 
from a trilinear (Yukawa) coupling. In particular, for the case of the
higgsinos, one finds $
%% \mathcal{L}^\text{SUSY} \supset 
\lambda \tilde h_u \tilde h_d S $. 
Since it allows for a scale invariant superpotential, as can be seen
from Eq.~(\ref{eq:NMSSM:W}), the NMSSM is in fact the simplest
supersymmetric generalization of the SM in which the SUSY breaking
scale is the only scale in the Lagrangian (notice that the 
EW scale originates exclusively from the SUSY breaking scale).

The scalar components of the singlet superfield mix with the neutral
scalar components of $\hat H_u$ and $\hat H_d$, leading to an enlarged
Higgs sector, which now comprises three scalars, $h_i^0$, and two
pseudoscalars, $a^0_i$. Likewise, the fermionic component of $\hat S$ 
(the singlino, $\chi^0_S$) mixes
with the neutral higgsinos and gauginos, so that now one has five
neutralinos. 
Depending on the regime considered, the new states can either decouple
from the rest of the spectrum (an ``effective''-MSSM scenario), be mixed
with the MSSM states, or even be the lightest Higgs and neutralino.  
One can thus have a richer
and more complex phenomenology, with a potential 
impact for low energy physics (e.g. flavour physics), dark
matter scenarios and searches at colliders.

\section{Higgs phenomenology in the NMSSM}
\begin{figure}[t!]
\begin{center}
\hspace*{-4mm}
\begin{tabular}{ccc}
\raisebox{2mm}{\epsfig{file=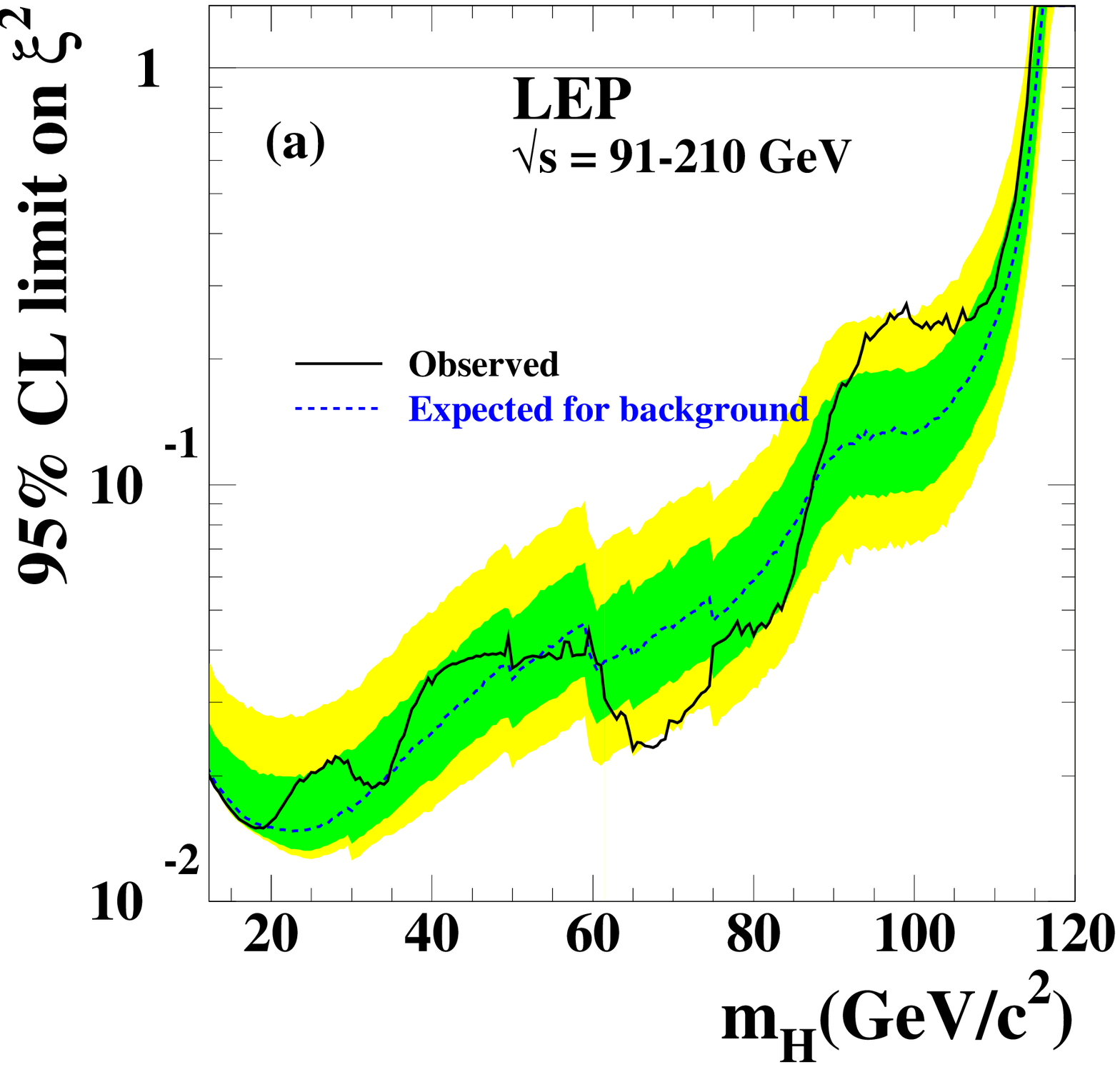, 
clip=, angle=0, width=50mm}}\hspace*{-4mm}&
\hspace*{-5mm}
\epsfig{file=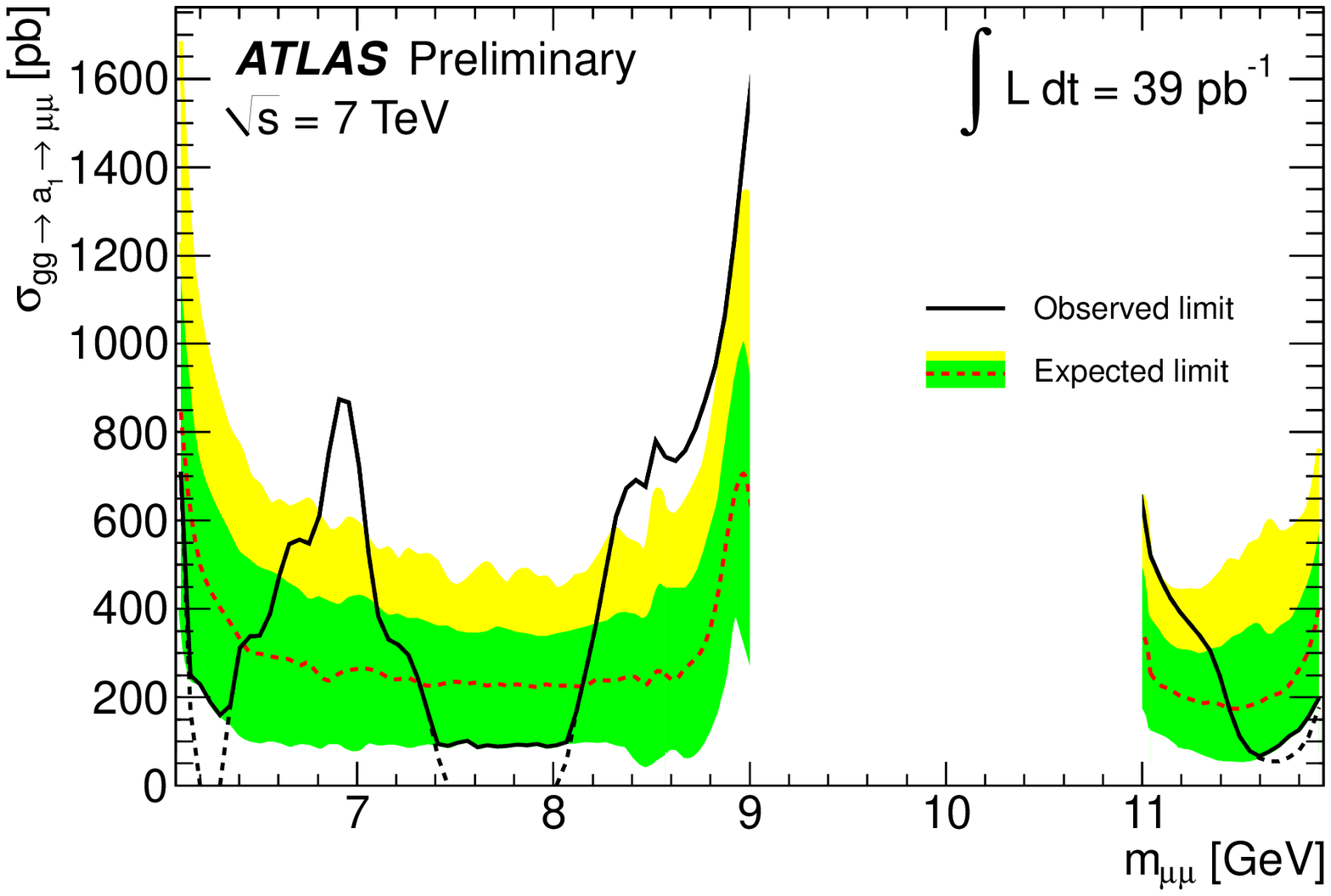, clip=, angle=0, width=73mm}
\hspace*{-3mm}&\hspace*{-5mm}
\raisebox{3mm}{\epsfig{file=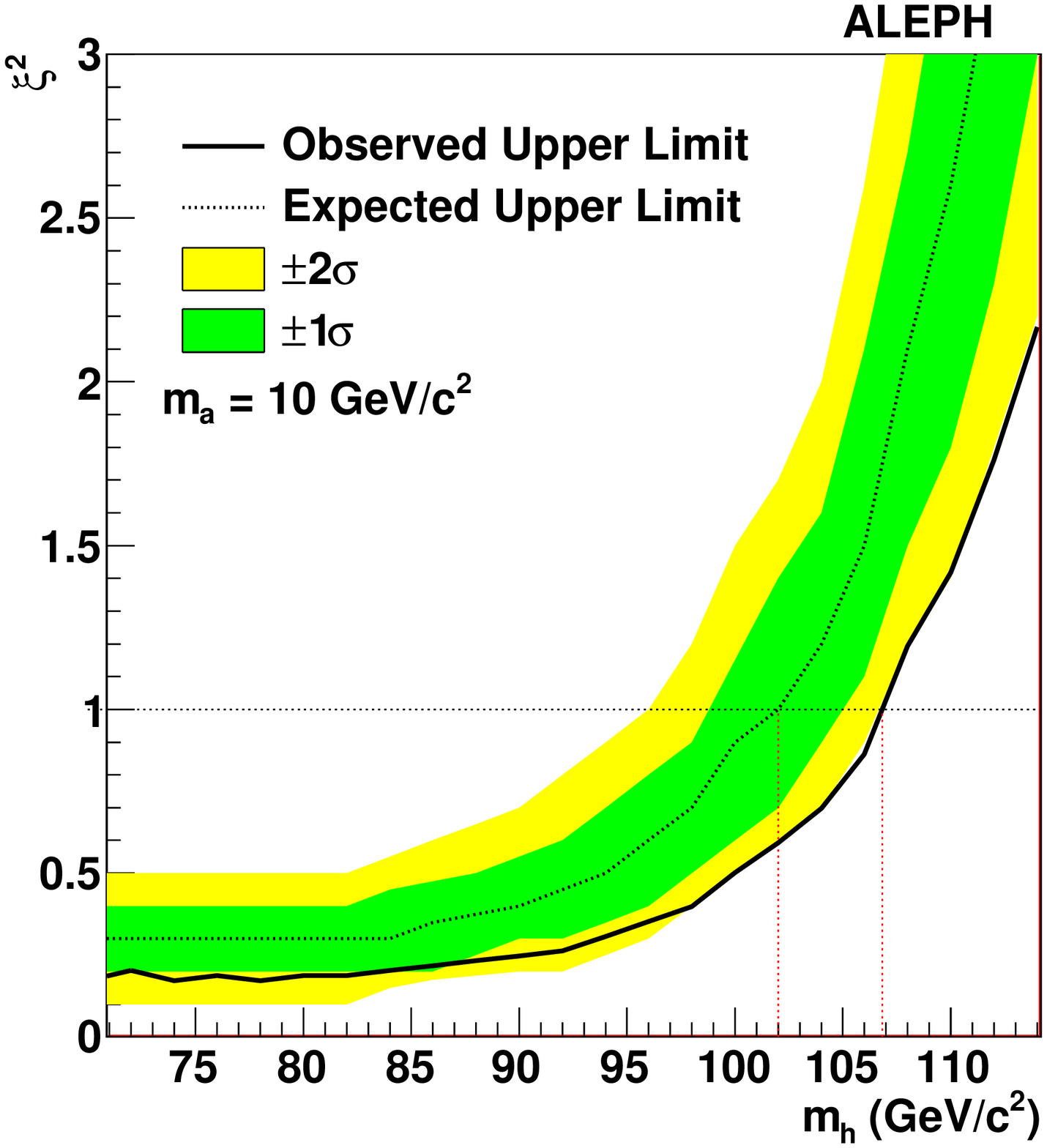, clip=, angle=0,
    width=47mm}}
\end{tabular}
\vspace*{-3mm}
\caption{From left to right: upper bound on $\xi^2$ $(={\xi^Z}^2)$ as
  a function of the scalar Higgs mass; upper limits on 
 $\sigma(gg \to a_1 \to \mu \mu)$ as a function of the dimuon invariant 
mass; upper bound on $\xi^2$ $(={\xi^A}^2)$ as a function of $m_H$, 
for $m_a=10$ GeV.
}
\label{fig:higgs.lep}
\end{center}
\end{figure}

When compared to the MSSM, 
the additional states and the new couplings of the NMSSM can
significantly alter the phenomenology of scalar and pseudoscalar
Higgs: 
in the NMSSM, both $h^0_1$ and $a_1^0$ can be very light, and still
comply with all collider and low-energy bounds. 
Firstly, if the lightest
scalar has a dominant singlet component, its reduced couplings to the
$Z$ boson ($\xi^Z \equiv g_{h_1ZZ}/g^\text{SM}_{HZZ}$) 
can be much smaller than in the MSSM~\cite{Ellwanger:1995ru}. As can be
seen from the left panel of Fig.~\ref{fig:higgs.lep},
depending on the value of $\xi^Z$ ($\xi=\xi^Z$), extremely light Higgs 
can still be
in agreement with the combined results from the four experiments at
LEP II. 

Higgs-to-Higgs decays are an extremely interesting and peculiar
feature of the NMSSM: in particular, in the presence of a light
(singlet-dominated) pseudoscalar, a SM-like $h_1^0$ 
($\xi^Z=1$) can have dominant
decays into a pair of light $a_1^0$ (thus reducing the $h^0_i \to b
\bar b, \, \tau^+ \tau^-$ branching ratios). 
Should this be the case,
then one can have $m_{h_1^0} \lesssim 114$ GeV, still in agreement with LEP 
data~\cite{lightA,Dermisek:2005gg}. 
Depending on the mass of $a_1^0$, it can decay to $b \bar b$, or into
a pair of charged leptons. There are presently strong constraints on
a light pseudoscalar, which we briefly summarise: for $m_{a_1^0}
\gtrsim 2 m_B$, LEP searches for $h^0_1 \to a_1^0 a_1^0 \to 4b$
strongly constrain $m_{h_1^0} \lesssim 100$ GeV; 
below the $b \bar b$
threshold, the most important constraints arise from $B$ and
$\Upsilon$ phenomenology, with
KLEO and BABAR severely constraining the regimes leading to $m_{a_1^0}
\lesssim 9$ GeV (the actual bounds depending on $X_d= \cos \theta_A
\tan \beta $, where $\cos \theta_A$ denotes the doublet-like component
of $a_1^0$)~\cite{B.upsilon}. 
This has been reinforced by recent ATLAS searches for 
a light pseudoscalar decaying into 
$\mu \mu$ pairs~\cite{Schumacher.EW2011}, 
as shown on the centre plot of Fig.~\ref{fig:higgs.lep}.
An NMSSM pseudoscalar with a mass $9 \text{ GeV} \lesssim
m_{a^0_1} \lesssim 10.1$ GeV satisfies all available
constraints and, if such a light state mixes with the
$\eta_b$ meson, it could also explain the observed
$\Upsilon(1s)-\eta_b(1s)$ hyperfine splitting~\cite{Domingo:2009tb}. 
On the right hand-side of Fig.~\ref{fig:higgs.lep}
we display the ALEPH bounds, under the assumption that 
a light pseudoscalar, $m_{a_1^0} \sim 10$ GeV, is present (in this case, 
${\xi^A}^2=\frac{\sigma (e^+ e^- \to Z h)}
{\sigma (e^+ e^- \to Z h)_\text{SM}} \times \text{BR}(h \to a a
)\times \text{BR}(a \to \tau^+ \tau^-)^2$).
 
As pointed out in~\cite{Dermisek:2010mg}, for $m_{a_1^0} \sim 10$ GeV, and if 
BR($a_1^0 \to \tau^+ \tau^-$)$\sim 80\%$ 
(i.e., ${{\xi^A}^2 \lesssim 0.5 - 0.6}$), LEP data allows 
a SM-like CP-even Higgs with  $m_{h_1^0} \sim 100$ GeV. This 
interesting NMSSM regime offers the possibility to reconcile 
LEP Higgs searches with EW precision measurements, the latter strongly
favouring $m_{H} \sim 100$ GeV. 

Finally, it is interesting to remark that the NMSSM also offers two
possible explanations for the slight excess of events ($\sim 2.3\,
\sigma$) observed at LEP for $m_H \sim 95 - 100$ GeV: either the
lightest Higgs has a non-vanishing singlet component~\cite{excess.singlet}
(leading to 
$\xi^Z \sim 0.4$, as can be infered from the left panel of 
Fig.~\ref{fig:higgs.lep}), or it is indeed SM-like, but dominantly
decays into a pair of light pseudoscalars~\cite{Dermisek:2010mg}, 
as discussed above. 

On the theoretical side, it is also relevant to notice
that in the NMSSM the mass of the SM-like Higgs can
be larger than in the MSSM~\cite{Ellwanger:2006rm}: 
for large values of $\lambda$ (but
$\lambda \lesssim 0.7$ to avoid a Landau pole below $M_\text{GUT}$) and
in the low $\tan \beta$ regime, one can have $m_{h_1^0} \sim 140$ GeV,
where $h^0_1$ has SM-like couplings to fermions and gauge bosons
($h^0_1 \sim H^\text{SM}$). 
In the limit where the lightest Higgs is singlet-like ($\xi^Z \to
0$), $h_2^0$ behaves as $H^\text{SM}$, its
mass being no larger than the above mentioned bound. However, in scenarios
of maximal mixing between doublet and singlet-like states, one can
have $m_{h_1^0} \gtrsim 110$ GeV and  $m_{h_2^0} \lesssim 162$ GeV,
stil in agreement with LEP data. Moreover,
should $h_1^0$ be singlet-like and decay unconventionally (e.g. 
$h_1^0 \to a_1^0 a_1^0 \to 4b$), then the upper bound on the mass of
the SM-like $h_2^0$ can be even further relaxed. 
For these regimes, Tevatron exclusion results~\cite{Tevatron.EW2011} 
already apply to part of the NMSSM parameter space (contrary to the MSSM 
case).

By relaxing the upper bound on the lightest Higgs boson, and
allowing for regimes where a light SM-like Higgs is still in agreement
with LEP bounds, the NMSSM also renders less severe the so-called 
``Higgs little fine tuning problem'' of the MSSM, 
which is related to the non-observation of a light Higgs state at LEP.
In the MSSM, the mass
of the lightest Higgs state is bounded from above: at
tree level $m_{h^0_1} \lesssim M_Z |\cos 2\beta|$, and while the
inclusion of radiative corrections allows to relax this bound, 
one still has $m_{h^0_1} \lesssim 130 -135$ GeV (the limits being 
model dependent, and for a 
sparticle spectrum no heavier than a few TeV). The
allowed interval for the mass of the lightest MSSM Higgs scalar
is thus considerably narrower than in the NMSSM.

\section{LHC search strategies}
Having an extended and more complex Higgs sector~\footnote{We will
  not discuss here the impact of an extended neutralino sector for
  sparticle production and decay at colliders.} does not imply that
detection of an NMSSM Higgs boson will be easier at the LHC. 
In the previous section we have seen that NMSSM Higgs might have
escaped LEP detection, either due to non-standard couplings to SM fermions
and gauge bosons, or in the presence of Higgs-to-Higgs decays. 
NMSSM searches at the LHC must strongly build upon LEP's lessons: the
new, distinctive features of the NMSSM, 
especially concerning the Higgs sector, 
must be taken into account in devising strategies, for instance 
for ATLAS and CMS.  
The different production processes, new intermediate states in cascade
decays, and unusual final-state configurations might require dedicated
studies and simulations.

If Higgs-to-Higgs decays are kinematically forbidden (or marginally
allowed, but with tiny branching ratios), then NMSSM Higgs searches 
can be carried
out as in the MSSM~\cite{Djouadi:2005gj}. 
Different couplings and new (loop) corrections
should be taken into account in a (re)-evaluation of the expected
production cross sections and decay rates. 
For some regimes, the Higgs sector can be more visible than in the
MSSM (e.g., as shown in~\cite{Moretti:2006sv}, up to 3 Higgs - $h_{1,2}^0$ and
$a_1^0$ - can be observed, from the decays into 2 photons). 
Recently, it was noticed that light
NMSSM Higgs, with a mass 80-100 GeV (in agreement with LEP constraints
due to a large singlet component) may have a BR($h_1^0 \to \gamma
\gamma$) considerably larger than
a SM-like Higgs of similar mass, $\sigma (gg \to h_1^ 0 \to
\gamma \gamma ) \sim 6 \times \sigma (gg \to H^ \text{SM} \to
\gamma \gamma )$, due to a reduced coupling to $b$
quarks~\cite{Ellwanger:2010nf}. 

In recent years, many efforts have been put forward to generalise the 
``No-lose'' theorem of the MSSM to the NMSSM: under the assumption that  
Higgs-to-Higgs decays are kinematically forbidden, it has been
established that at least one of the NMSSM Higgs bosons can be detected
at the LHC with 600 fb$^{-1}$ of integrated luminosity~\cite{Ellwanger:2001iw}.

However, not only Higgs-to-Higgs decays can occur in large regions
of the NMSSM parameter space, but they also constitute one of the most
interesting features of this model.
If these decays are indeed present, Higgs searches at the LHC (and at
the Tevatron) can be considerably more complicated, and many new channels have
been considered, for the different $m_{a_1^0}$ regimes. Here we
briefly comment on some dedicated strategies for regions in
parameter space where the dominant decays of (light) Higgs are $h_1^ 0
\to a_1^0 a_1^0$, and $a_1^0 \to \tau \tau$. (For $m_{a_1^0}$ above
the $b \bar b$ threshold, see for instance~\cite{a.to.4tau}.)
In general, it can be quite challenging to identify the four leptons 
in these decay modes, and
final states containing as much as 8 neutrinos imply signatures of
large missing energy. SM backgrounds will also be important (heavy
flavour jets, vector boson and light jets, $\Upsilon$ production, etc.).

The $h_1^0 \to a_1^0a_1^0 \to 4\tau$ channel, with the taus decaying
into muons and jets, has been analysed in~\cite{Belyaev:2008gj}, resorting
to both Higgs-strahlung (triggering
on leptonic decays of $W^\pm$), and vector boson fusion (triggering on two same
sign non-isolated muons). While the latter may yield a larger number
of events, the former can lead to very clean, almost background free
signals, so that in both cases there is a significant potential
for discovery. In regions where $a_1^0 \to \mu \mu$, the $2\mu$
($4\mu$) invariant mass allows a direct estimation of $m_{a_1^0}$
($m_{h_1^0}$); furthermore, the extremely small background allows to
rely on direct $gg$ and $b\bar b$ fusion
for Higgs production (instead of the subdominant vector
boson fusion)~\cite{Belyaev:2010ka}.
If the lightest Higgs is produced via central exclusive production,
$pp \to h_1^0 \to p + h_1^0 +p$ (with $h_1^0 \to a_1^0a_1^0 \to
4\tau$), the prospects for observing such an NMSSM Higgs at the LHC are
good, and one could determine $m_{h_1^0}$ and $m_{a_1^0}$ on an
event-to-event basis. However, this would require installing forward
detectors to measure the final state protons~\cite{Forshaw:2007ra}.
Finally, for regimes of very low $\tan \beta$ ($\tan \beta \lesssim
2$), most of LHC (and Tevatron) discovery prospects must be
reconsidered: in such regimes for $\tan \beta$, 
BR($a_1^0\to \tau^+ \tau^-$) becomes increasingly
reduced (accompanied by an increase of BR($a_1^0 \to gg + c \bar c$)), 
so that the light
pseudoscalar easily evades both ALEPH and meson physics constraints 
(due to small
${\xi^A}^2$ and $X_d$). However, this also implies that searches using
the $a_1^0 \to \tau \tau$ and $a_1^0 \to \mu \mu$ modes will be more
difficult. Nevertheless, dedicated searches at the LHC and Tevatron
include direct detection of $a_1^0$ in $g g \to a_1^0 \to \mu \mu$
channel (as wel as in the other channels mentioned 
before)~\cite{Dermisek:2010mg}.

Light singlet-like Higgs are very difficult to detect (due to the
smallness of their couplings). It has been noticed that in this case 
the process $pp \to h_1^0 + \text{ resolved jet } \to \tau^+\tau^- +$ jet
(via gluon fusion) could allow for LHC detection with $\sqrt{s}=14$
TeV~\cite{Belyaev:2010bc}.

In NMSSM scenarios with a light doublet-like CP-odd Higgs boson, 
the charged Higgs can be 
lighter than the top quark, dominantly decaying as $h^\pm
\to a_1^0 W^\pm$. The search for subleading $a_1^0$ decay modes 
(into a pair of muons) 
could provide evidence for the charged Higgs, or even a discovery,
with early LHC data~\cite{Dermisek:2010af}.
Other channels, which are absent in the MSSM, and that deserve further
investigation are, for example, $gg \to a_2^0 \to h^\pm W^\mp$ (where
the $a_2^0$ has an important singlet component)~\cite{Mahmoudi:2010xp}. 

It is important to re-emphasise
that the discovery of MSSM-like Higgs and neutralinos does 
not necessarily establish that the MSSM is indeed at work:
disentangling the NMSSM from the MSSM might be 
challenging, especially in regimes where the new states decouple
and/or in the absence of a singlino LSP. In this case 
additional studies might be required, and unravelling the nature of 
the SUSY model will strongly depend on the precision of the experimental
data.

\section{Implications for Dark Matter}
Due to the differences in the neutralino and Higgs sectors of the
NMSSM, one can have dark matter scenarios that are very distinct from
the MSSM. Depending on the regions of the parameter space,
the LSP can be singlino-like (or have an important singlino
component). 
The additional scalar and pseudoscalar Higgs bosons 
can have in impact on the processes leading to LSP annihilation, so
that the correct relic density can be obtained in large regions of the
parameter space~\cite{DM.NMSSM}: the extra states can
offer rapid annihilation via new s-channel resonances, and if light,
new final states can be kinemmatically open (e.g. annihilation into $Z
h_1^0$, $h_1^0 h_1^0$, $h_1^0 a_1^0$ and $a_1^0 a_1^0$). For instance,
nearly pure binos can efficiently annihilate via $h_1^0$ resonances
into a pair of light $a_1^0 a_1^0$. Provided there is a small
higgsino component, a singlino LSP can also rapidly annihilate 
via the latter process and co-annihilations with heavier neutralinos,
or with a nearly degenerate NLSP, are also possible. A singlino LSP
can also be instrumental in recovering MSSM scenarios with a charged
LSP (e.g., the lightest stau for $m_0 \ll M_{1/2}$ in the constrained
MSSM). 
 
Dark matter detection prospects can also be significantly
different~\footnote{This topic was also addressed in the talks of A.
  Goudelis and T. Delahaye.}. 
As discussed in~\cite{light.DM} light NMSSM neutralinos (with
a mass below the MSSM lower bound) 
may have an elastic scattering cross section on nucleons allowing
to explain recent direct detection results (DAMA/LIBRA, CoGeNT or
CDMS), provided that the spectrum contains light scalar and
pseudoscalar Higgs.

\section{A simple and predictive model: the constrained NMSSM}
Assuming that supersymmetry is spontaneously broken in a hidden
sector, and that the mediation of SUSY breaking to the observable
sector occurs via flavour blind interactions (as is the case of
minimal supergravity models), all soft SUSY breaking terms will be
universal at some very large scale (e.g., $M_\text{GUT}$). The scale
invariant NMSSM with universal soft breaking terms is denoted the
fully constrained NMSSM (cNMSSM)~\cite{cNMSSM}, 
and is one of the most appealing SUSY
extensions of the SM, both for its simplicity and predictivity. 

Other than the gauge and quark/lepton Yukawa couplings, the 
Lagrangian of the cNMSSM depends on five parameters - 
$m_0^2$, $M_{1/2}$, $A_0$, $\lambda$ and $\kappa$ -, the correct EW
symmetry breaking reducing the parameter space from five to four
degrees of freedom. 
However, phenomenological arguments strongly constrain the parameter
space, as we proceed to briefly explain. 

\begin{figure}[t!]
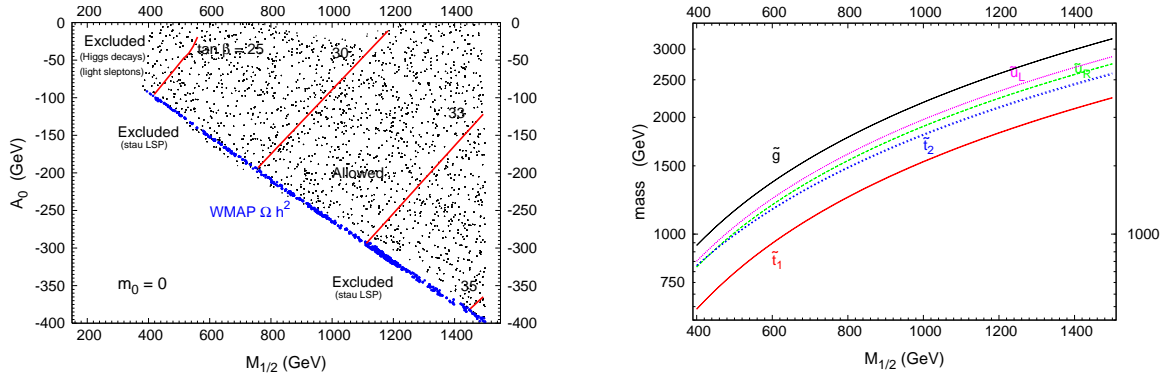

\begin{center}
\hspace*{-0mm}
\begin{tabular}{cc}
\epsfig{file=figs/Plot.M12A0scan.epsi, clip=, angle=270, width=70mm}
\hspace*{3mm}&
\hspace*{3mm}
\epsfig{file=figs/Plot.3006.m12.maxm0.phenoGQL.epsi, 
clip=, angle=270, width=70mm}
\end{tabular}
\caption{On the left, cNMSSM parameter space:
  experimentally allowed regions (scatter points) and imposing the
  correct DM relic density (blue). On the right, cNMSSM gluino and squark
  spectrum as a function of $M_{1/2}$.}
\label{fig:cNMSSM}
\end{center}
\end{figure}

In order to generate a non-vanishing singlet VEV
(as required by the lower bound on the effective $\mu$-term 
($|\mu| \gtrsim 100$ GeV), the singlet soft breaking mass $m_S^2$ must
not be too large. Since $m_{_S}$ is hardly renormalised between the GUT
and the EW scales, its value at $M_\text{GUT}$, given
by $m_0$, must also be small (compatible with $m_0 \sim 0$). While in the
cMSSM a regime where $m_0 \lesssim 1/5 \,M_{1/2}$ would lead to a
charged LSP (the lightest stau), in the cNMSSM 
the additional singlino-like neutralino can be lighter than $\tilde
\tau_1$, so that a viable dark matter candidate can be recovered for
very small or even vanishing values of $m_0$. An efficient reduction of
the LSP abundance can only be achieved via co-annihilations with the
stau NLSP, requiring nearly degenerate LSP and NLSP ($m_{\tilde
\tau_1} -m_{\chi_S^0} \sim $ few GeV), which implies that $A_0 \sim -1/4\,
M_{1/2}$ (and furthermore $m_0 \leq 1/10 \,M_{1/2}$). Under such a
regime for the soft breaking parameters, LEP constraints on the Higgs
sector imply that $\lambda$ must be also very small, $\lambda \lesssim
0.02$. Provided $\lambda$ is not excessively small ($\lambda \gtrsim
10^{-5}$, to allow for co-annihilation), the resulting phenomenology is
largely independent of its exact value. Thus, as depicted on the left
panel of Fig.~\ref{fig:cNMSSM}, the parameter space
of the fully constrained NMSSM is essentially determined by $M_{1/2}$
($\tan \beta$, no longer a free parameter, is quite large, $\tan \beta
> 25$). Collider constraints lead to $M_{1/2} \gtrsim 500$ GeV, while
the requirement that SUSY contributions account for the discrepancy of
the measured muon anomalous magnetic moment with respect to the SM
prediction favours $M_{1/2} \lesssim 1$ TeV~\cite{Domingo:2008bb}. 

Concerning the Higgs sector of the cNMSSM, and for increasing values
of $M_{1/2}$, the lightest state can be singlet-like, a
doublet-singlet mixture and, for large $M_{1/2}$, SM-like (the actual
cross-over range depending on the value of $m_0$). The lightest
pseudoscalar (always heavier than $h_{1,2}^0$) is singlet-like, while 
$h_3^0$, $a_2^0$ and $h^\pm$ are significantly heavier and 
nearly degenerate. Interestingly, just below the singlet-doublet 
cross-over for $h_{1,2}^0$, the cNMSSM can actually account for the two 
LEP ``excesses'', with a singlet-like $h_{1}^0$ with mass around 100
GeV and a SM-like $h_{2}^0$ around 117 GeV. 
The cNMSSM strongly interacting sparticle spectrum, displayed on the
right hand-side of Fig.~\ref{fig:cNMSSM}, is quite heavy (typically 
$m_{\tilde g, \tilde q} \gtrsim 1$ TeV), with the gluino always
heavier than all squarks (and sleptons). As seen from
Fig.~\ref{fig:cNMSSM}, the measurement
of one sparticle mass (or mass difference) would allow to 
predict quite accurately the remaining sparticle spectrum.

Having a
singlino LSP, nearly degenerate with the NLSP, leaves a striking
imprint on cNMSSM decay chains: due to the weakly coupled singlino-like LSP, all
sparticle branching ratios into $\chi_{_S}^0$ are tiny, and thus sparticles first
decay into the stau NLSP. As an example, the simplest squark cascades
typically are $\tilde q \to q \chi_2^0 \to q \tilde \tau_1 \tau \to
q \tau \tau \chi_S^0$. Hence, practically all cascade
decays will go via $\tilde \tau_1$, leading to two $\tau$'s per
decaying squark. 
For very small $\lambda$, or a very small NLSP-LSP mass
difference, the stau lifetime can be so large that its decay vertices
are visibly displaced, $\mathcal{O}$(mm - cm), a 
``smoking-gun'' for the cNMSSM. 
All the above features should in principle allow 
to discriminate the cNMSSM from most realisations of the
MSSM.

Another very appealing feature of the cNMSSM is that it can be easily
ruled out. Detection of a singlino LSP relies on its non-singlet
component, which is  $\mathcal{O}(\lambda)$; hence direct detection
(LSP-nucleon) cross
sections are extremely small, and indirect detection of the products
of LSP annihilation also appears impossible. Thus, the direct or
indirect detection of a weakly interacting massive particle allows to
exclude the cNMSSM. 

The prospects for cNMSSM discovery at the LHC have been discussed
in~\cite{Ellwanger:2010es}. 
The dominant sparticle production modes are squark-gluino
and squark pair production. Regarding the SM-like Higgs $h_{1,2}^0$, the 
most relevant production processes will be gluon-gluon and vector boson
fusion, $gg \to$ Higgs and $qq \to qq+$Higgs, with the Higgs decaying 
into two photons
(possibly $\tau^+ \tau^-$ in the vector boson fusion
process). The heavier non-singlet Higgs can be observed in associated
production with $b \bar b$ pairs while, apart from the ``cross-over''
region, the singlet-like Higgs states are generally inaccessible. 
Dedicated cNMSSM cuts suggest that for the LHC operating at $\sqrt s=$
14 TeV, and for an integrated luminosity of 1 fb$^{-1}$, the
signal-to-background ratio already allows for the discovery of the
cNMSSM in the lower $M_{1/2}$ regime, while more luminosity will be
required in the case of a heavier spectrum. 
Furthermore, the cNMSSM can be distinguished from
the MSSM in the stau co-annihilation region.

\section{Outlook}

The NMSSM is a very interesting SUSY extension of the SM, solving in
an elegant way the ``$\mu$-problem'' of the MSSM, and rendering its ``Higgs 
little fine tuning problem'' less severe. Since it allows for a scale
invariant superpotential, the NMSSM is the simplest supersymmetric
model in which the SUSY breaking scale is the only scale in the
Lagrangian. 

The extended Higgs and neutralino sectors of the NMSSM have an impact regarding
low-energy observables (such as $B$ physics), dark matter prospects and 
collider phenomenology. Concerning the latter, the NMSSM allows to
accommodate LEP constraints easier than the MSSM. In
particular, the upper bound on the mass of the SM-like Higgs boson 
is relaxed, and the lightest
Higgs scalar and pseudoscalar can be quite light (either due to an
important singlet component, or to unconventional decays, such as $h
\to a a$).  Unconventional Higgs decay scenarios require 
dedicated studies and simulations. At present 
many studies are under way to ensure that at least one NMSSM 
Higgs will be observed at the LHC. The absence of a ``No-lose''
theorem should be kept in mind: a non-discovery of a Higgs boson at the
LHC (potentially excluding scenarios as the cMSSM) could be a signal
of the NMSSM.

The cNMSSM is perhaps one of the
most simple and yet most predictive supersymmetric extensions of the
SM since, in addition to all the appealing features of the NMSSM, its
phenomenology is essentially described by one parameter,
$M_{1/2}$. The cNMSSM predicts a heavy sparticle spectrum,
with a $\tilde \tau_1$ appearing in all cascades, leading to a
singlino-like LSP. The model can
be discovered at the LHC, and be easilly ruled out by dark matter
detection.

\end{document}